\documentstyle[prl,aps]{revtex} 
\begin{document}
\twocolumn[\hsize\textwidth\columnwidth\hsize\csname
@twocolumnfalse\endcsname 
\title{Neutron Star Structure and the Neutron Radius of $^{208}$Pb}
\author{C.~J.~Horowitz
\footnote{e-mail:  charlie@iucf.indiana.edu} 
}
\address{Nuclear Theory Center and Dept. of Physics, Indiana
University, Bloomington, IN 47405}
\author{J.  Piekarewicz
\footnote{e-mail: jorgep@scri.fsu.edu}
}
\address{Department of Physics Florida State University, 
Tallahassee, FL 32306}
\date{\today} 
\maketitle 
\begin{abstract}
We study relationships between the neutron-rich skin of a heavy
nucleus and the properties of neutron-star crusts. Relativistic
effective field theories with a thicker neutron skin in $^{208}$Pb
have a larger electron fraction and a lower liquid-to-solid transition
density for neutron-rich matter. These properties are determined by
the density dependence of the symmetry energy which we vary by adding
nonlinear couplings between isoscalar and isovector mesons. An
accurate measurement of the neutron radius in $^{208}$Pb---via parity
violating electron scattering---may have important implications for
the structure of the crust of neutron stars.
\end{abstract}
\vskip2.0pc]

It is an extrapolation of 18 orders of magnitude from the neutron
radius of a heavy nucleus---such as $^{208}$Pb with 
$R_{n}\!\approx\! 5.5$~fm---to the approximately 10 km radius of a neutron 
star. Yet both radii depend on our incomplete knowledge of the 
equation of state of neutron-rich matter. Therefore, an accurate 
measurement of the neutron radius in $^{208}$Pb may have important 
implications for some neutron star properties.

Heavy nuclei are expected to have a neutron-rich skin. This important
feature of nuclear structure arises because of the large neutron
excess and because the Coulomb barrier reduces the proton density at
the surface. The thickness of the neutron skin depends on the pressure
of neutron-rich matter: the greater the pressure, the thicker the skin
as neutrons are pushed out against surface tension. The same pressure
supports a neutron star against gravity~\cite{lat}. Thus models with thicker 
neutron skins often produce neutron stars with larger radii.

Neutron stars are expected to have a solid crust of nonuniform
neutron-rich matter above a liquid mantle. The phase transition from
solid to liquid depends on the properties of neutron-rich
matter. Indeed, a high pressure implies a rapid rise of the energy
with density making it energetically unfavorable to separate uniform
matter into regions of high and low densities.  Thus a high pressure
typically implies a low transition density from a solid crust to a
liquid mantle. This suggests an inverse relationship: the thicker the
neutron-rich skin of a heavy nucleus, the thinner the solid crust of 
a neutron star.

In this letter we study possible ``data-to-data'' relations between
the neutron-rich skin of a heavy nucleus and the crust of a neutron
star. These relations may impact neutron star observables. Indeed, 
properties of the crust are important for models of glitches in the 
rotational period of pulsars~\cite{glitch,glitch2}, for the shape 
and gravitational radiation of non-spherical rotating
stars~\cite{grav} and for neutron-star cooling~\cite{brem}. Note 
that the skin of a heavy nucleus and the crust of a neutron star 
are composed of the same material: neutron-rich matter at similar
densities.

The Parity Radius Experiment (PREX) at the Jefferson Laboratory aims
to measure the neutron radius in $^{208}$Pb via parity violating
electron scattering~\cite{prex,bigpaper}.  Parity violation is
sensitive to the neutron density because the $Z^0$ boson couples
primarily to neutrons.  The result of this purely electroweak
experiment could be both accurate and model independent. In contrast,
all previous measurements of bulk neutron densities used hadronic
probes that suffer from controversial uncertainties in the reaction
mechanism (see for example Ref.~\cite{ray}). PREX should provide a
unique observational constraint on the thickness of the neutron skin
in a heavy nucleus. In this letter we explore some of the implications
of this measurement on the structure of neutron stars.

Microscopic calculations of the energy of 
neutron matter constrain both the neutron skin in $^{208}$Pb 
and the crust of a neutron star; see for example Ref.~\cite{vj}.  
However, these calculations of infinite neutron matter are not directly 
tested by observable properties of finite nuclei such as their 
charge densities or binding energies. Moreover, nonrelativistic 
calculations of symmetric nuclear matter have not succeeded in 
predicting the saturation density. It thus becomes necessary 
to fit some properties of a three-body force in order to 
reproduce nuclear saturation. Indeed, the properties of 
$A\!=\!8$ pure neutron drops calculated in Ref.~\cite{drops} 
may depend on the three-nucleon force used. Thus we feel 
that it is important to distinguish direct finite-nucleus 
measurements---such as PREX---from theoretical neutron-matter
``observables'' based solely on calculations.  Indeed, PREX 
may provide an important test of these calculations~\cite{brown}.

We start with a relativistic effective field theory~\cite{horst} that
provides a simple description of finite nuclei and a Lorentz covariant
extrapolation for the equation of state of dense neutron-rich matter.
The theory has an isoscalar-scalar $\phi$ (sigma) meson field and
three vector fields: an isoscalar $V$ (omega), an isovector {\bf $b$}
(rho), and the photon $A$. We now supplement the Lagrangian with  
new nonlinear sigma-rho and omega-rho couplings. These couplings 
allow us to change the density dependence of the symmetry energy 
which changes both the thickness of the neutron skin in $^{208}$Pb 
and the neutron-star crust.

The interacting Lagrangian density is given by~\cite{horst},
\begin{eqnarray}
&&
{\cal L}_{\rm int} =
\bar\psi \left[g_{\rm s}\phi   \!-\! 
         \left(g_{\rm v}V_\mu  \!+\!
    \frac{g_{\rho}}{2}{\mbox{\boldmath $\tau$}}\cdot{\bf b}_{\mu} 
                               \!+\!    
    \frac{e}{2}(1\!+\!\tau_{3})A_{\mu}\right)\gamma^{\mu}
         \right]\psi \nonumber \\
                   && - 
    \frac{\kappa}{3!} (g_{\rm s}\phi)^3 \!-\!
    \frac{\lambda}{4!}(g_{\rm s}\phi)^4 \!+\!
    \frac{\zeta}{4!}   g_{\rm v}^4(V_{\mu}V^\mu)^2 \!+\!
    \frac{\xi}{4!}     g_{\rho}^4({\bf b}_{\mu}\cdot {\bf b}^\mu)^2
    \nonumber \\
                   && +
    g_{\rho}^{2}\,{\bf b}_{\mu}\cdot{\bf b}^{\mu}
    \left[\Lambda_{4}g_{\rm s}^{2}\phi^2 +
            \Lambda_{\rm v} g_{\rm v}^2V_{\mu}V^\mu\right] \;.
\end{eqnarray}

We consider a number of different parameter sets. First, we note that
the nonlinear rho coupling $\xi$ will modify the density dependence of
the rho mean field and this could change the neutron-skin
thickness. However, unless $\xi$ is made very large, this term was
found to have a small effect~\cite{horst}. Thus for simplicity, 
we set $\xi\!\equiv\!0$ in all our parameter sets. Second, note
that we could have added a cubic sigma-rho interaction of the form:
${\cal L}_{3}=M \Lambda_{3}(g_{\rm s}\phi) (g_{\rho}^{2}\,{\bf
b}_{\mu}\cdot{\bf b}^{\mu})$. A nonzero $\Lambda_3$ does change the
thickness of the neutron skin in $^{208}$Pb---but at the expense of 
a change in the proton density. Therefore, we set 
$\Lambda_3\!\equiv\!0$ and focus exclusively on $\Lambda_4$ and 
$\Lambda_{\rm v}$.

We start with the original NL3 parameter set of Lalazissis, K\"onig, 
and Ring~\cite{NL3}. (Note that a small adjustment of the $NN\rho$ 
coupling constant was needed to fit the symmetry energy of nuclear
matter at a Fermi momentum of $k_{F}\!=\!1.15 {\rm fm}^{-1}$; see
text below). The NL3 set has 
$\zeta\!=\!\Lambda_4\!=\!\Lambda_{\rm v}\!=\!0$ and provides a good 
fit to the ground-state properties of many nuclei. In this model 
symmetric nuclear matter saturates at $k_F\!=\!1.30$~fm$^{-1}$ with 
a binding energy per nucleon of $E/A\!=\!-16.25$~MeV and an 
incompressibility of $K\!=\!271$~MeV. All other parameter sets 
considered here have been fixed to the same saturation properties.

We now add the new nonlinear couplings $\Lambda_4$ and/or
$\Lambda_{\rm v}$ between the isoscalar and the isovector mesons. At
the same time we adjust the strength of the $NN\rho$ coupling constant
($g_\rho$) from its NL3 value to maintain the symmetry energy of
nuclear matter unchanged (see text below). Note that 
neither $\Lambda_4$ nor $\Lambda_{\rm v}$ affect the properties of 
symmetric nuclear matter since ${\bf b}_\mu\!\equiv\!0$. Hence, 
the saturation properties remain unchanged. Our goal is to change 
the neutron density and the neutron-skin thickness in $^{208}$Pb 
while making very small changes to the proton density which is well 
constrained by the measured charge density~\cite{sick}. The two 
new couplings ($\Lambda_4$ and $\Lambda_{\rm v}$) change the skin 
thickness in $^{208}$Pb by similar amounts. Yet they have different 
high-density limits\cite{horst}. For $\Lambda_{\rm v}\!=\!0$ 
the symmetry energy 
is proportional to the baryon density $\rho$ in the limit of very 
high density, while it only grows like $\rho^{1/3}$ for nonzero 
$\Lambda_{\rm v}$. This can produce noticeable differences in 
neutron star radii, as we show below.

Let us start with $\Lambda_4\!=\!0$. For a given omega-rho coupling
$\Lambda_{\rm v}$ we readjust only the $NN\rho$ coupling constant
$g_\rho$ in order to keep an average symmetry energy fixed. The 
symmetry energy at saturation density is not well constrained by 
the binding energy of nuclei. However, some average of the symmetry 
energy at full density and the surface energy is constrained 
by binding energies. As a simple approximation we evaluate the 
symmetry energy not at full density $k_F\!\approx\!1.30$~fm$^{-1}$ 
but rather at an average density corresponding to 
$k_F\!=\!1.15$~fm$^{-1}$. Thus, all our parameter sets have a 
symmetry energy of $23.50$~MeV at $k_F\!=\!1.15$~fm$^{-1}$. Note 
that the original NL3 parameter set predicts a symmetry energy of 
37.4 MeV at full saturation density and close to 23.50 MeV at 
$k_F\!=\!1.15$ fm$^{-1}$~\cite{NL3}. This simple procedure produces 
a nearly constant binding energy per nucleon for $^{208}$Pb as 
$\Lambda_{\rm v}$ is changed, as can be seen in Table~\ref{tableone}. 
Moreover, Table~\ref{tableone} shows that increasing $\Lambda_{\rm v}$ 
reduces the neutron-skin thickness significantly---while maintaining 
the proton radius nearly constant. In the following we plot our 
results for a range of $\Lambda_{\rm v}$ values for which the 
proton radius is within 0.01 fm of its $\Lambda_{\rm v}\!=\!0$ 
value.

To study the solid crust of a neutron star we make a simple 
random-phase-approximation (RPA) calculation of the transition 
density below which uniform neutron-rich matter becomes 
unstable against small amplitude density fluctuations. This 
provides a lower bound to the true transition 
density~\cite{bound}. We start with the longitudinal dielectric 
function $\epsilon_L$, as defined in Eq.~(68) of 
Ref.~\cite{wehrberger}, evaluated at an energy transfer 
$q_0\!=\!0$ and at an arbitrary momentum transfer $q$. That is,
\begin{equation}
\epsilon_L(q_0\!=\!0,q)=\det\left(1-D_L \Pi_L\right)\;.
\end{equation}
Here $\Pi_L$ is a longitudinal polarization matrix describing 
particle-hole excitations of a uniform system of protons, 
neutrons, and electrons in beta equilibrium, as given in 
Eq.~(56) of Ref.~\cite{wehrberger}. The matrix $D_L$, describing 
meson and photon propagation, follows from Eq.~(57) of 
Ref.~\cite{wehrberger}---but includes additional terms to account 
for the nonlinear nature of the meson self-couplings~\cite{future}. 
We estimate the transition density $\rho_{c}$ by computing  
the largest density at which $\epsilon_L(0,q)\!<\!0$ for any 
given $q$.

%%%%%%%%%%%%%%%%%%%%%%%%%%%%%%%%%%%%%%%%%%%%%%%%%%%%%%%
\begin{table}[h]
\caption{Results for the NL3 Parameter Set. The binding 
energy per particle in $^{208}$Pb is B.E., $R_p$ is the 
proton and $R_n-R_p$ is the difference between neutron 
and proton radii in Pb. (Note that we do not include 
center of mass corrections.)  Finally, $\rho_c$ is our 
estimate for the transition density of neutron-rich 
matter from a nonuniform to uniform phase.}
\begin{tabular}{lccccc}
$\Lambda_v$& $g_\rho^2$ & B.E. (MeV) & $R_p$ (fm) 
& $R_n-R_p$ (fm) & $\rho_c$ (fm$^{-3}$) \\
0.0 	& 79.6 	&  7.85 & 5.460 & 0.280 & 0.052 \\
0.005   & 84.9  &  7.86 & 5.461 & 0.265 & 0.056 \\
0.01    & 90.9  &  7.87 & 5.462 & 0.251 & 0.061 \\
0.015   & 97.9  &  7.88 & 5.463 & 0.237 & 0.067 \\
0.02    & 106.0 &  7.88 & 5.466 & 0.223 & 0.075 \\
0.025   & 115.6 &  7.89 & 5.469 & 0.209 & 0.081 
\label{tableone}
\end{tabular}
\end{table}
%%%%%%%%%%%%%%%%%%%%%%%%%%%%%%%%%%%%%%%%%%%%%%%%%%%%%%%

\vbox to 2.5in{\vss\hbox to 8in{\hss {\includegraphics{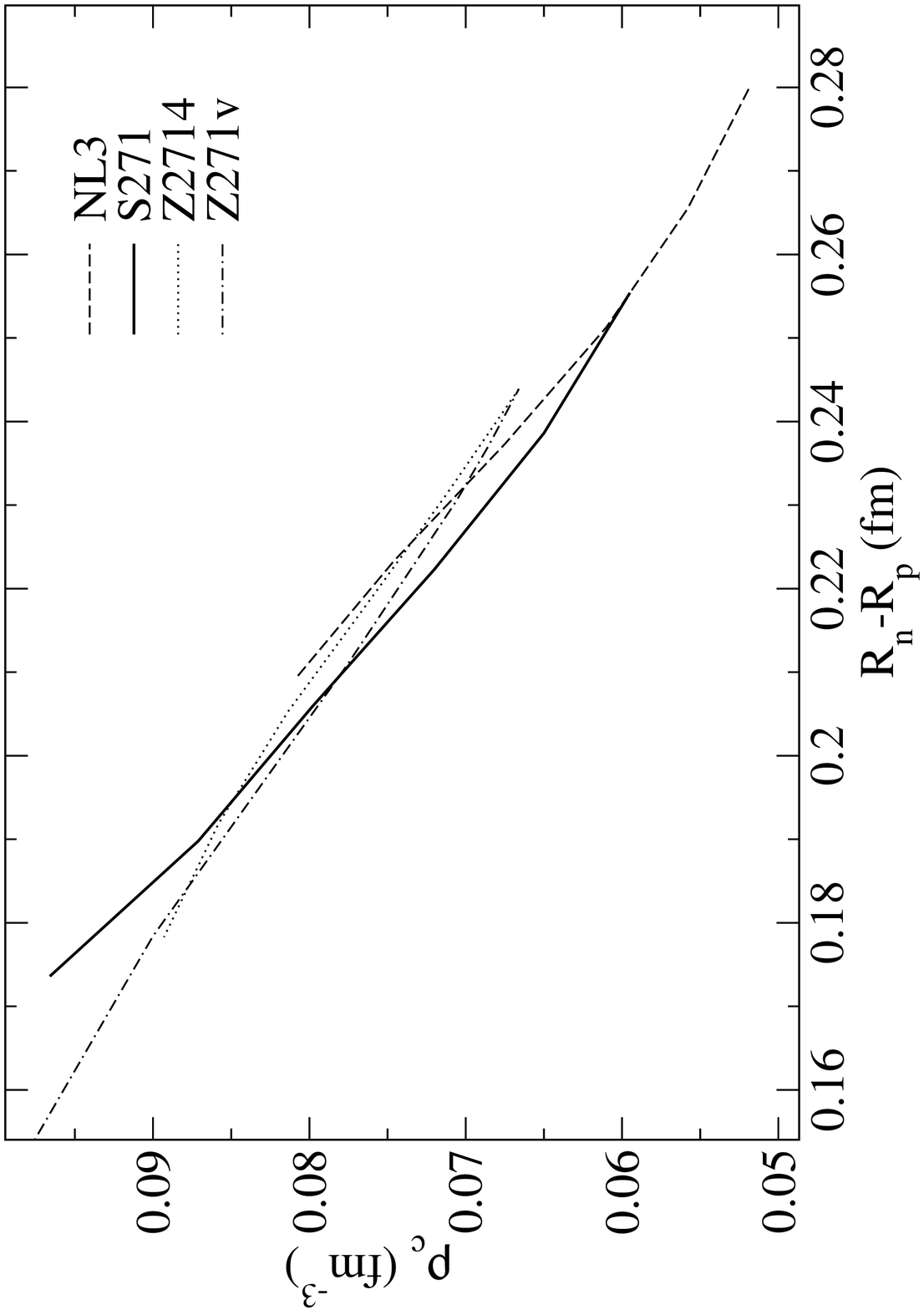}}\hss}}
\nobreak {\noindent\narrower{{\bf FIG.~1}. Estimate of the transition
density from nonuniform to uniform neutron-rich matter versus 
neutron-minus-proton radius in $^{208}$Pb. The curves are for
the four parameter sets described in the text.}}
\vskip 8pt

In Fig. 1 we display the transition density for various parameter 
sets (see Table~\ref{tabletwo}) as a function of the predicted 
difference in the root-mean-square neutron and proton radii 
$R_n\!-\!R_p$ in $^{208}$Pb. The curves are parameterized by different 
values of $\Lambda_{\rm v}$, as shown in Table~\ref{tableone}. 
The NL3 parameter set saturates nuclear matter with a relatively 
small value of the nucleon effective mass: 
$M^*\!\equiv\!M\!-\!g_s\phi\!=\!0.59M$. The parameter set S271 
saturates nuclear matter as NL3 but with $M^*\!=\!0.70M$. This set 
also has $\zeta\!=\!0$. The two remaining curves in the figure are 
for parameter sets having $\zeta\!=\!0.06$ and both saturate nuclear 
matter with $M^*\!=\!0.80M$. (Set Z271v has a nonzero 
$\Lambda_{\rm v}$ while set Z2714 uses a nonzero 
$\Lambda_4$). Note that the scalar mass $m_{\rm s}$ for parameter 
sets S271, Z271v, and Z2714 is adjusted to reproduce the proton 
radius in $^{208}$Pb as computed with NL3. Figure 1 displays a clear 
inverse correlation between the transition density and the neutron-skin 
thickness $R_n\!-\!R_p$. The transition density expressed in 
fm$^{-3}$ is about,
\begin{equation}
\rho_{c}\approx 0.16 - 0.39\,(R_n-R_p) \;,
\end{equation}
with the skin thickness expressed in fm. Moreover, this correlation
seems to be insensitive to $M^*$ or to using $\Lambda_4$ or 
$\Lambda_{\rm v}$ to change $R_n\!-\!R_p$. These results suggest 
that a measurement of the neutron radius in $^{208}$Pb will 
provide considerable information on the transition density.

\begin{table}
\caption{Model parameters used in the calculations. The 
parameter $\kappa$ and the scalar ($m_{\rm s}$) and vector 
($m_{\rm v}$) masses are given in MeV. The nucleon and rho 
masses are kept fixed at $M\!=\!939$ and $m_\rho\!=\!763$ MeV, 
respectively.}

\begin{tabular}{lccccccc}
Model & $g_{\rm s}^2$ & $g_{\rm v}^2$ & $\kappa$ & 
        $\lambda$ & $\zeta$ & $m_{\rm s}$ & $m_{\rm v}$ \\ 
NL3  & 104.387 & 165.585 & 3.860 & $-$0.01591 & 0 & 508.194 & 782.5 \\ 
S271 &  85.992 & 116.766 & 6.68  & $-$0.01578 & 0 & 505     & 783   \\ 
Z271 &  53.785 &  70.669 & 6.17  &    0.15634 & 0.06 & 465  & 783   
\label{tabletwo}
\end{tabular}
\end{table}

Note that Fig. 1 only shows our results. Yet all other calculations 
that we are aware of also give consistent results. For example, the
nonrelativistic microscopic equation of state of Friedman and
Pandharipande has a transition density of $\rho_c=0.096$\ fm$^{-3}$\
according to Lorenz et al.~\cite{vj}.  For this equation of state
Brown finds $R_n-R_p=0.16\pm 0.02$\ fm~\cite{brown}. These numbers
are in excellent agreement with Eq. (3).

Brown also finds a linear relation between $R_n-R_p$\ and the
derivative of the energy of neutron matter versus density $dE/d\rho$\
evaluated at $\rho=0.1$\ fm$^{-3}$~\cite{brown}.  He considers a large
variety of nonrelativistic Skyrme interactions.  Our results for
$dE/d\rho$ versus $R_n-R_p$ are completely consistent.  We expect
these common $dE/d\rho$ values to give similar $\rho_c$ values
consistent with Eq. (3) for these Skyrme interactions.  This is
because $dE/d\rho$ is related to the pressure while $\rho_c$ depends
on the density dependence of the pressure.

Finally, for the relativistic interaction TM1 of Sugahara and
Toki~\cite{TM1}, we calculate from Eq. (2) $\rho_c\approx 0.059$\
fm$^{-3}$ and $R_n-R_p=0.27$ fm.  Again, the numbers are again in 
good agreement with Eq. (3).

In Fig.~2 we show the electron fraction per baryon $Y_e$ versus 
density for uniform neutron-rich matter in beta equilibrium. We 
include results only for the S271 parameter set as all other sets 
yield similar results. The different curves are for different 
values of $\Lambda_{\rm v}$ which predict the indicated 
$R_n\!-\!R_p$ values. The curves start near the transition 
densities displayed in Fig.~1. The electron fraction $Y_e$ is
determined by the symmetry energy while $R_n\!-\!R_p$ is sensitive 
to the density dependence of the symmetry energy. Therefore a 
measurement of $R_n\!-\!R_p$ constrains the growth of $Y_e$ with 
density. If $R_n\!-\!R_p$ is greater than about 0.24 fm, 
$Y_e$ becomes large enough to allow the direct URCA 
process~\cite{urca} to cool down a 1.4 solar mass neutron star.

\vbox to 2.95in{\vss\hbox to 8in{\hss {\includegraphics{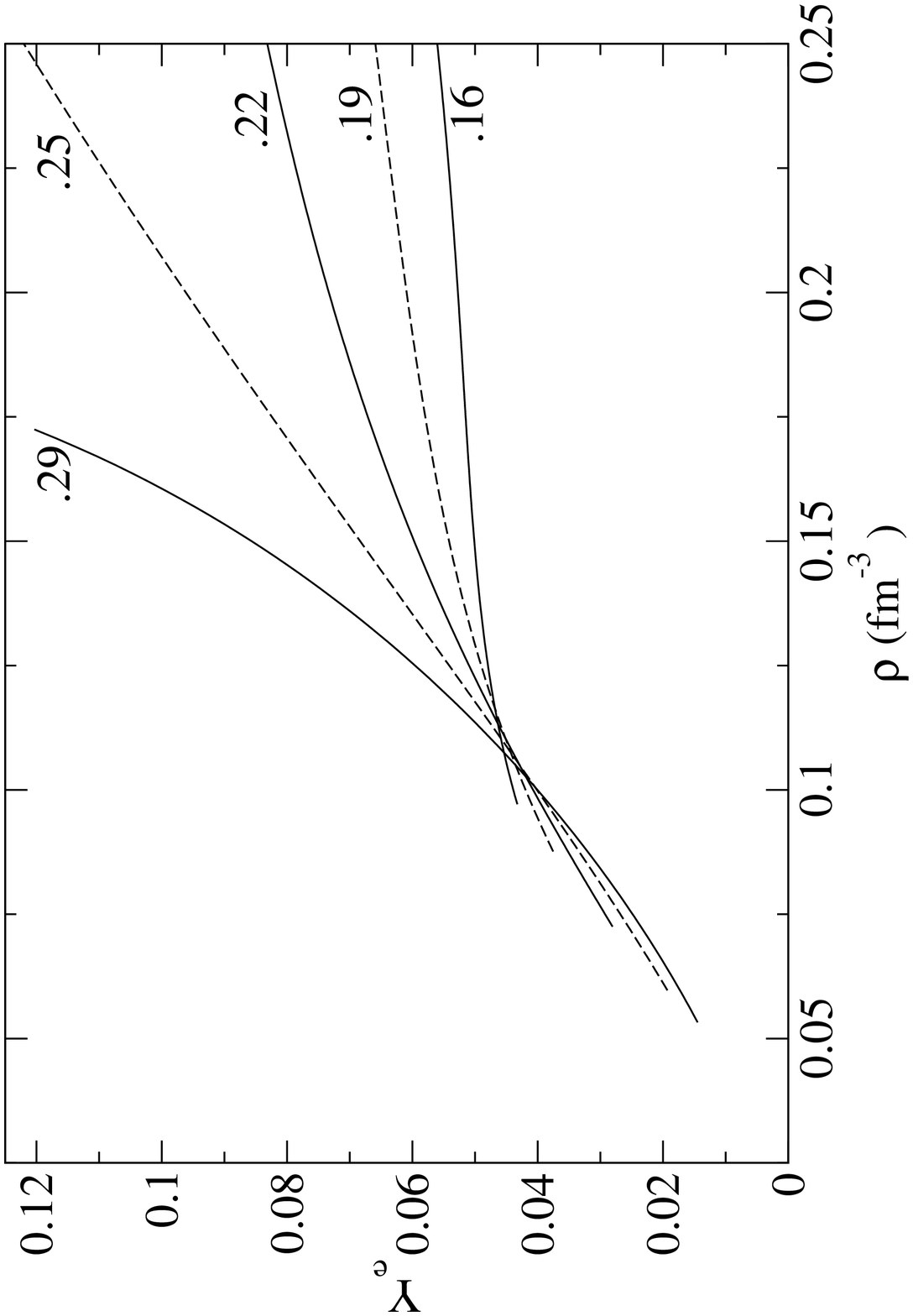}}\hss}}
\nobreak {\noindent\narrower{{\bf FIG.~2}. Electron fraction $Y_e$
versus baryon density for uniform neutron-rich matter in beta
equilibrium using the S271 parameter set (other sets yield similar
results). The curves are for different values of $\Lambda_{\rm v}$ 
that predict the indicated values of $R_n-R_p$ for $^{208}$Pb, in fm.}}
\vskip 10pt

We now consider the radius $R$ of a 1.4 solar mass neutron star.  For a
given parameter set, $R$ increases with $R_n\!-\!R_p$.  However, as one
changes the parameter set to increase $M^*$ or $\zeta$ the equation of
state becomes softer at high density.  As a result $R$ decreases for
fixed $R_n\!-\!R_p$.  Also, the high density equation of state is
softer with $\Lambda_{\rm v}$ than with $\Lambda_4$ so Z271v gives
slightly smaller stars than parameter set Z2714.  We conclude that $R$
is not uniquely constrained by a measurement of the neutron-skin
thickness because $R_n\!-\!R_p$ only depends on the equation of state
at normal and lower densities while $R$ is also sensitive to the
equation of state at higher densities. Yet one may be able to combine
separate measurements of $R_n\!-\!R_p$ and $R$ to obtain considerable
information about the equation of state at low and high densities.
For example, if $R_n\!-\!R_p$ is relatively large while $R$ is small
this could suggest a phase transition.  A large $R_n\!-\!R_p$ implies
that the low density equation of state is stiff while a small $R$
suggests that the high density equation of state is soft.  The
transition from stiff to soft could be accompanied by a phase
transition.

\vbox to 2.95in{\vss\hbox to 8in{\hss {\includegraphics{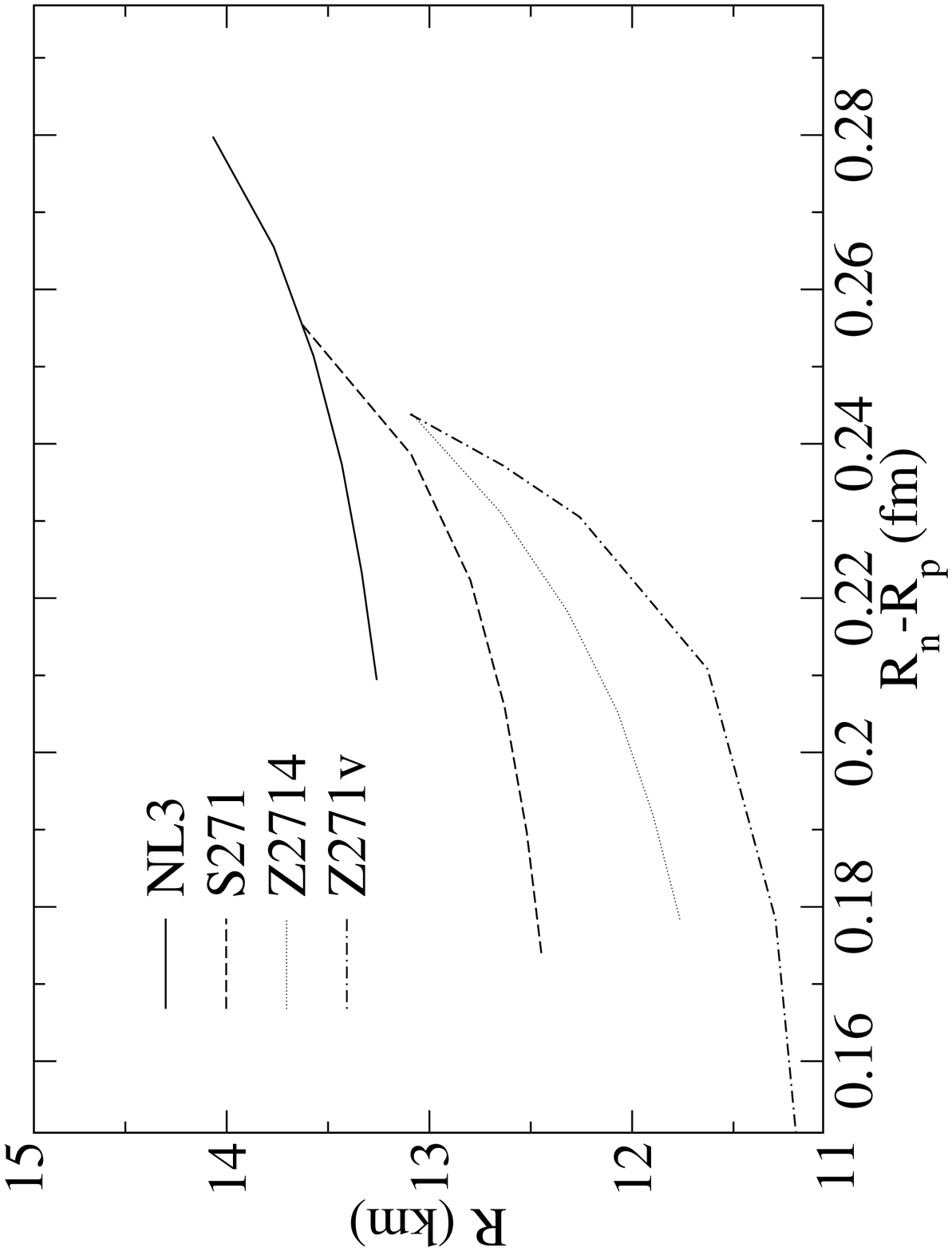}}\hss}}
\nobreak {\noindent\narrower{{\bf FIG.~3}. Radius of a 1.4 solar 
mass neutron star versus neutron-minus-proton radius in 
$^{208}$Pb for the four parameter sets described in the text.}}
\vskip 10pt

In conclusion:
1) It is possible to fit nuclear observables---such as
charge densities, binding energies, and single particle 
spectra---with effective field theories that predict a
range of neutron-skin thicknesses. This can be done by adding 
nonlinear couplings between isoscalar and isovector meson fields 
or, in general, by adding interactions that modify the density 
dependence of the symmetry energy. We conclude that the neutron-skin 
thickness is not tightly constrained by these observables. Yet a 
measurement of the skin thickness will constrain the density 
dependence of the symmetry energy.

2) The density dependence of the symmetry energy is adjustable 
in our relativistic effective field models while still reproducing 
nuclear-matter properties and other ground-state observables. Indeed,
our models can provide a Lorentz-covariant extrapolation for the high 
density equation of state with a symmetry energy that rises slower 
with density relative to earlier relativistic mean-field models.

3) The electron fraction $Y_e$ of neutron-rich matter in beta 
equilibrium is correlated with the neutron-skin thickness in 
$^{208}$Pb. The thicker the neutron skin the faster $Y_e$ rises with 
density. In our models a neutron-skin thickness of the order of
0.24~fm or larger suggests that $Y_e$ will become large enough to 
allow a direct URCA process to cool down a $1.4 M_{\odot}$ neutron 
star.

4) We have found an inverse correlation between the 
neutron-skin thickness and the density of a phase transition 
from nonuniform to uniform neutron-rich matter. In our models 
the transition density obeys the approximate relation: 
$\rho_c\approx 0.16 -0.39\,(R_n\!-\!R_p)$, with $R_n\!-\!R_p$ 
in fm and $\rho_c$ in fm$^{-3}$. This suggests that a neutron 
skin measurement in $^{208}$Pb can provide important information 
on the thickness and other properties of the crust of a neutron 
star.

5) Microscopic calculations of the energy of neutron matter 
constrain the density dependence of the symmetry energy and hence 
the neutron-skin thickness in $^{208}$Pb. Therefore a neutron skin 
measurement may provide an important observational check on such 
calculations that is not provided by other observables. Moreover, 
a neutron-skin measurement may constrain three-body forces in 
neutron-rich matter.

\smallskip
We acknowledge useful discussions with A. Brown, R. Furnstahl,
J. Lattimer, S. Reddy and B. Serot.  We thank A. Brown for sharing
results prior to publication.  This work was supported in part by DOE
grants DE-FG02-87ER40365 and DE-FG05-92ER40750.

\end{document}